\documentclass[reprint,aps,pra,amsmath,amssymb,superscriptaddress,floatfix]{revtex4-2}

\usepackage{textcomp}
\usepackage{bm}
\usepackage{amsmath}
\usepackage{graphicx}

\usepackage{color}
\usepackage{xspace}% correct spacing after formulas from macros
\usepackage{bm}% bold math
\usepackage{xcolor}% color

\begin{document}

\title{Slope of the upper critical field at $T_{c}$ in two-band superconductors
with non-magnetic disorder: $s_{++}$ superconductivity in $\textrm{Ba}_{1-x}\textrm{K}_{x}\textrm{Fe}_{2}\textrm{As}_{2}$}

\author{R.~Prozorov}
\email[Corresponding author: ]{prozorov@ameslab.gov}
\affiliation{Ames National Laboratory, Ames, IA 50011, U.S.A.}
\affiliation{Department of Physics \& Astronomy, Iowa State University, Ames, IA 50011, U.S.A.}

\author{V.~G.~Kogan}
\affiliation{Ames National Laboratory, Ames, IA 50011, U.S.A.}

\author{M.~Ko\'{n}czykowski}
\affiliation{Laboratoire des Solides Irradi\'{e}s, CEA/DRF/IRAMIS, \'{E}cole Polytechnique,
CNRS, Institut Polytechnique de Paris, F-91128 Palaiseau, France}

\author{M.~A.~Tanatar}
\affiliation{Ames National Laboratory, Ames, IA 50011, U.S.A.}
\affiliation{Department of Physics \& Astronomy, Iowa State University, Ames, IA 50011, U.S.A.}

\date{17 June 2023}

\begin{abstract}
A recent theory of the disorder-dependent slope of the upper critical field, $H_{c2}$, at the superconducting transition temperature, $T_{c}$, is extended to multiband superconductors aiming at iron-based superconductors,
considering two constant gaps of different magnitude and, potentially, different signs. The result shows that there is only a narrow domain inside the $s_{\pm}$ pairing state where the slope increases with the increase of transport (non-magnetic) scattering rate, $P$. In most phase space, the slope should decrease in an $s_{\pm}$ state and increase in the $s_{++}$ pairing state. The experiment shows that in an archetypal iron-based superconductor, $\textrm{Ba}_{1-x}\textrm{K}_{x}\textrm{Fe}_{2}\textrm{As}_{2}$ (BaK122), non-magnetic disorder induced by electron irradiation increases the slope $S$ across the superconducting ``dome,'' at different $x$. This implies that $\textrm{Ba}_{1-x}\textrm{K}_{x}\textrm{Fe}_{2}\textrm{As}_{2}$ is likely an $s_{++}$ superconductor with two (or more) gaps of different magnitudes. This work reopens a decade-long discussion of the nature of the superconducting order parameter in iron pnictides.
\end{abstract}
\maketitle

\section{Introduction}

Soon after discovering iron-based superconductors (IBS), it became obvious that although practically all these materials are fully gapped, measured thermodynamic quantities did not follow the clean single isotropic gap predictions. Instead, it was found that natural and deliberately introduced non-magnetic disorder is pair-breaking \cite{Gordon2010,Kogan2009}. Even in the clean limit (quite achievable in iron pnictides due to the extremely short coherence length of 2-3 nm \cite{Meier2016}), adding non-magnetic disorder led to significant suppression of the superconducting transition temperature, $T_{c}$, violating the Anderson theorem \cite{Anderson1959JPCS}. The natural solution was the so-called $s_{\pm}$ pairing due to spin fluctuations promoted by nesting \cite{Mazin2008}. This concept was generalized to include states where nesting does not play a pivotal role, but still, the order parameter changed its sign between some of the Fermi surface sheets, see for reviews Refs.\cite{Mazin2009PhysicaC,Johnston2010AP,Paglione2010review,Hirschfeld2011,Chubukov2012,Chubukov2015}.
The experimental suppression of $T_{c}$ by disorder is then well described by the extension of the Abrikosov - Gor'kov (AG) theory of magnetic impurities in conventional isotropic $s_{++}$ superconductors \cite{Abrikosov1960} to anisotropic order parameter, including $s_{\pm}$ \cite{Openov1998,Kogan2009a,Timmons2020}.

Further intense research showed that the results of transport and thermodynamic measurements could be explained with both anisotropic $s_{\pm}$ or $s_{++}$ pairing due to spin and orbital fluctuations,
respectively \cite{Kontani2012}. Angle-resolved photoemission spectroscopy (ARPES) revealed complex anisotropic doping-dependent electronic band structure
and multiple energy gaps with similarities but also differences between IBS ``families'' \cite{Ding2008,Borisenko2010,Okazaki2012,Umezawa2012,Mou2016,Shimojima2017}. With anisotropic and different in magnitude gaps on realistic Fermi surface pockets, some with distinct 3D character, with \cite{Mou2016} or without \cite{Borisenko2010} nesting undermines the original highly symmetric picture of an $s_{\pm}$ pairing \cite{Mazin2010}. At the same time, the $\ensuremath{s_{++}}$ pairing was not completely dismissed  \cite{Umezawa2012,Kontani2012,Shestakov2018}. Moreover, a crossover from $s_{\pm}$ to $\ensuremath{s_{++}}$ was predicted \cite{EfremovPRB2011} and even observed \cite{Torsello2019}. The problem is that, so far, there are no phase-sensitive experiments for any iron pnictides yet. For iron chalcogenides, in particular, Fe(Se,Te), phase-sensitive quasiparticle interference measurements have been interpreted in favor of an $s_{\pm}$ pairing  \cite{Hanaguri2010,Sprau2017}. Unfortunately, the 122 compounds do not cleave in the way needed for such experiments \cite{Hanaguri2023}.

In most IBS, including the subject of this study, $\textrm{Ba}_{1-x}\textrm{K}_{x}\textrm{Fe}_{2}\textrm{As}_{2}$, $T_{c}\left(x\right)$ shows a dome-like variation. However, the slope of the upper critical field, $S\equiv\partial H_{c2}/\partial T$ estimated at $T_{c}$ in pristine samples, appears to be a simple linear function of $T_{c}$ as expected from the BCS theory in the clean limit \cite{Kogan2023}. The slope $S$ is considered a very useful quantity because it is used to estimate the zero-temperature value of the upper critical field, $H_{c2}\left(0\right)$, which is mostly inaccessible experimentally in IBS. One can estimate the coherence length from $H_{c2}\left(0\right)$ and discuss possible Pauli limiting.
Unfortunately, the HW (Helfand-Werthamer) \cite{Helfand1966,Helfand1964} or more often quoted WHH (Werthamer-Helfand-Hohenberg) \cite{Werthamer1966}
theory only dealt with isotropic $s-$wave superconductors. Their theory and obtained coefficients do not apply to anisotropic superconductivity
considered first by father and son Pokrovsky \cite{Pokrovsky1996} and recently cast in a more accessible form in Ref.\cite{Kogan2023}.
Here we extend the latter approach even further to the two-band superconductivity and apply the conclusions to analyze the experimental data obtained in electron-irradiated hole-doped BaK122 crystals.

Even a brief literature survey finds numerous reports of the upper critical field slope at $T_{c}$ as a function of disorder introduced by various means in various materials. The overall experimental picture
is quite clear - superconductors with line nodes show decreasing slope, whereas those without nodes show increasing $dH_{c2}/dT$. The original two-gap $s_{++}$ superconductor, MgB$_{2}$, shows an increase of
the slope $S$ with increasing residual resistivity \cite{Shabanova2008}. A dramatic decrease of $T_{c}$, almost a complete suppression was reported in Mn substituted MgB$_{2}$, yet the slope of $H_{c2}$
remained practically unchanged \cite{Rogacki2006}. In another proven two-gap superconductor, V$_{3}$Si \cite{Cho2022}, a pronounced increase
of the slope was found after neutron irradiation \cite{Meier-Hirmer1982}. On the nodal side, we have high$-T_{c}$ cuprates, - hole-doped YBa$_{2}$Cu$_{3}$O$_{7-x}$
\cite{Antonov2020} and electron-doped (NdCe)$_{2}$CuO$_{4+y}$ \cite{Charikova2009}.
For quite some time, electron-doped cuprates were contrasted to the hole-doped ones as fully gapped. In YBCO, a clear $T-$linear variation of the London penetration depth, $\lambda(T)$, meant the existence of line nodes, but in electron-doped superconductors, it took seven more years before a similar but weaker claim, based on the quadratic behavior of $\lambda(T)$, characteristic of dirty nodal superconductor, was made \cite{Prozorov2000}. Now we can say that this is confirmed by the measured decrease of the $H_{c2}$ slope \cite{Charikova2009}.

In IBS, the transition temperature decreases with the non-magnetic disorder. For example, in electron-doped $\textrm{Ba}(\textrm{Fe}_{1-x}\textrm{Co}_{x})_{2}\textrm{As}_{2}$
where disorder was introduced by ball-milling \cite{Tokuta2019}. Similar effect is found in irradiated hole-doped $\textrm{Ba}_{1-x}\textrm{K}_{x}\textrm{Fe}_{2}\textrm{As}_{2}$ IBS after fast neutron irradiation \cite{Karkin2014} and 2.5 MeV electron irradiation (this work). With regard to the slope $S$, a steady decrease was found in isovalently substituted $\textrm{BaFe}_{2}(\textrm{As}_{1-x}\textrm{P}_{x})_{2}$.
In a heavily electron-irradiated sample, $T_{c}$ was suppressed below 10 K, while the slope $S$ monotonically decreased with irradiation dose  \cite{Konczykowski2023}. However, this particular IBS is unique among 122 compounds; it is nodal \cite{AsP122Science2012} and the observed decrease is consistent with our results. Another IBS,
NdFeAs(O,F), showed a monotonic increase of $dH_{c2}/dT$ upon irradiation with alpha particles \cite{Tarantini2018}. Strain and doping were shown to increase $S$ in $\textrm{Ba}_{1-x}\textrm{K}_{x}\textrm{Fe}_{2}\textrm{As}_{2}$
\cite{Tarantini2011}.

In this work, we extend the single band theory of Ref.\cite{Kogan2023}
to a two-band scenario needed to describe the iron-based superconductors.
Analyzing the data collected on electron-irradiated BaK122, we conclude that barring some very special and unrealistic set of parameters, the increasing slope $S$ puts them into an $s_{++}$ domain. This is an unorthodox conclusion, and we hope our work will stimulate further studies.

\section{The slope of $H_{c2}$ at $T_{c}$}

Let us assume often used separation of variables in the order parameter \cite{Markowitz1963}, $\varDelta\left(\mathbf{k},T\right)=\Psi\left(T\right)\Omega\left(\mathbf{k}\right)$, where the angular part is normalized via its Fermi surface average
$\left\langle \Omega^{2}\right\rangle _{FS}=1$ \cite{Pokrovskii1961,Kogan2002} and the $\Psi\left(T\right)$ function is obtained from the self-consistency equation \cite{Kogan2021}. We call this an $\Omega-$model. Without magnetic scattering, the critical temperature of materials with some
$\Omega\left(\mathbf{k}\right)$ is given by \cite{Openov1998,Kogan2009a}:
\begin{eqnarray}
\ln t_{c}+(1-\langle\Omega\rangle^{2})\left[\psi\left(\frac{P/t_{c}+1}{2}\right)-\psi\left(\frac{1}{2}\right)\right]=0\qquad\label{eq:tc}
\end{eqnarray}
\noindent where $t_{c}=T_{c}/T_{c0}$ and $P$ is the dimensionless transport (non-magnetic) scattering parameter (rate). Obviously, the Anderson theorem is readily recovered for isotropic $s-$wave superconductors, where $\Omega=1$.

\begin{widetext}
According to (son and father) Pokrovsky \cite{Pokrovsky1996}, cast
in the present form in Ref.\cite{Kogan2023}, the slope of the upper critical field along the $c-$axis of a uniaxial superconductor is given by,
\begin{eqnarray}
\frac{\partial H_{c2}}{\partial T}\Big|_{T_{c}}=-\frac{8\pi\phi_{0}T_{c0}}{\hbar^{2}}\frac{t_{c}\left[1+(1-\langle\Omega\rangle^{2})\psi^{\prime}\left(\frac{1}{2}+\frac{P}{2t_{c}}\right)\right]}{h_{3,0}\langle\Omega^{2}v_{a}^{2}\rangle+2(P/2t_{c})h_{3,1}\langle\Omega\rangle\langle\Omega v_{a}^{2}\rangle+(P/2t_{c})^{2}h_{3,2}\langle\Omega\rangle^{2}\langle v_{a}^{2}\rangle}\,.\qquad\label{eq:slope-general}
\end{eqnarray}
\noindent where, $\psi$ are digamma functions, $v_{a}$ are in-plane Fermi velocity and all coefficients $h_{\mu,\nu}(x)$ are evaluated at $x=P/2t_{c}$.
These coefficients are:
\begin{eqnarray}
h_{3,0} & = & -\frac{1}{2}\psi^{\prime\prime}\left(\frac{1}{2}+x\right)\,,\qquad h_{3,1}=\frac{1}{x^{3}}\Big[\psi\left(\frac{1}{2}+x\right)-\psi\left(\frac{1}{2}\right)-x\psi^{\prime}\left(\frac{1}{2}+x\right)++\frac{x^{2}}{2}\psi^{\prime\prime}\left(\frac{1}{2}+x\right)\Big]\,,\nonumber \\
h_{3,2} & = & \frac{1}{2x^{4}}\Big\{\pi^{2}x-6\left[\psi\left(\frac{1}{2}+x\right)-\psi\left(\frac{1}{2}\right)\right]+4x\psi^{\prime}\left(\frac{1}{2}+x\right)-x^{2}\psi^{\prime\prime}\left(\frac{1}{2}+x\right)\Big\}\,.\label{eq:h-i-j}
\end{eqnarray}
\end{widetext}

\begin{figure}[tb]
\includegraphics[width=8.5cm]{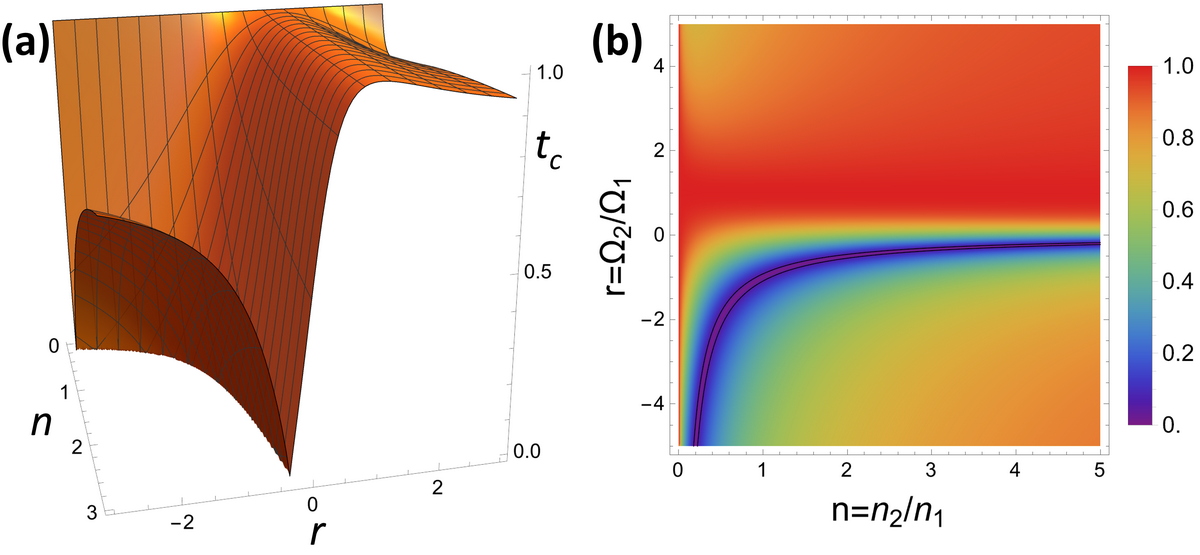}\caption{Superconducting transition temperature normalized by its pristine
value, $t_{c}=T_{c}/T_{c0}$, plotted as function of $n=n_{2}/n_{1}$
and $r=\Omega_{2}/\Omega_{1}$. The deep trench on the 3D plot,
corresponding to the middle of the blue $nr+1=0$ curve is the line
where $t_{c}$$\rightarrow0$, according to Eq.\ref{eq:tc} with Eq.\ref{eq:om}.}
\label{fig1:tc}
\end{figure}

\section{The two-band model}

Since we are dealing here with  multiband superconductors,
we need at least a two-band model. The full-blown two-band theory contains
many microscopic details, effects of which on a macroscopic features like the
slopes of $H_{c2}$ are not easy to track. Hence, we employ the simplest "minimum 
2-band model" within which we can calculate the slopes. Such two-band $\Omega-$model was first introduced for MgB$_2$   to explain the temperature-dependent anisotropy of London
penetration depth \cite{Kogan2002}. Here we adopt the same approach to analyze the slope of $H_{c2}$ at $T_c$. 

Let us consider two order parameters, $\Omega_{1}$ and $\Omega_{2}$ residing on two bands with the partial densities of states (DOS) at the Fermi level, $n_{1,2}=N_{1,2}/N$, where $N=N_{1}+N_{2}$ is the total density of states so that $n_{1}+n_{2}=1$. The normalization equation reads  \cite{Kogan2002}:
\begin{equation}
\left\langle \Omega^{2}\right\rangle =n_{1}\left\langle \Omega_{1}^{2}\right\rangle +n_{2}\left\langle \Omega_{2}^{2}\right\rangle =1
\label{eq:om-norm}
\end{equation}
Simplifying Eq.~\ref{eq:om-norm} even further, consider $\Omega_{i}$ to be constants (that can have different signs, though!) and introduce the ratio, $r=\Omega_{2}/\Omega_{1}$
and the ratio of the partial densities of states, $n=n_{2}/n_{1}$, we obtain,
\begin{eqnarray}
\ensuremath{\Omega_{1}^{2}=\frac{1}{n_{1}+n_{2}r^{2}}}\,,\qquad\ensuremath{\Omega_{2}^{2}=\frac{r^{2}}{n_{1}+n_{2}r^{2}}}\,.\label{eq:Omegas}
\end{eqnarray}

Therefore, the average, needed for Eq.\,(\ref{eq:tc}) is,

\begin{equation}
\left\langle \Omega\right\rangle ^{2}=\frac{(nr+1)^{2}}{(n+1)\left(nr^{2}+1\right)}\label{eq:OmegaSquared}
\end{equation}
and Eq.\,(\ref{eq:tc}) for the transition temperature now reads,

\begin{equation}
\ln t_{c}+\frac{n\left(r-1\right)^{2}}{(n+1)\left(nr^{2}+1\right)}\left[\psi\left(\frac{P}{2t_{c}}+\frac{1}{2}\right)-\psi\left(\frac{1}{2}\right)\right]=0\label{eq:KoganTerm}
\end{equation}

\begin{figure}[t]
\includegraphics[width=8.5cm]{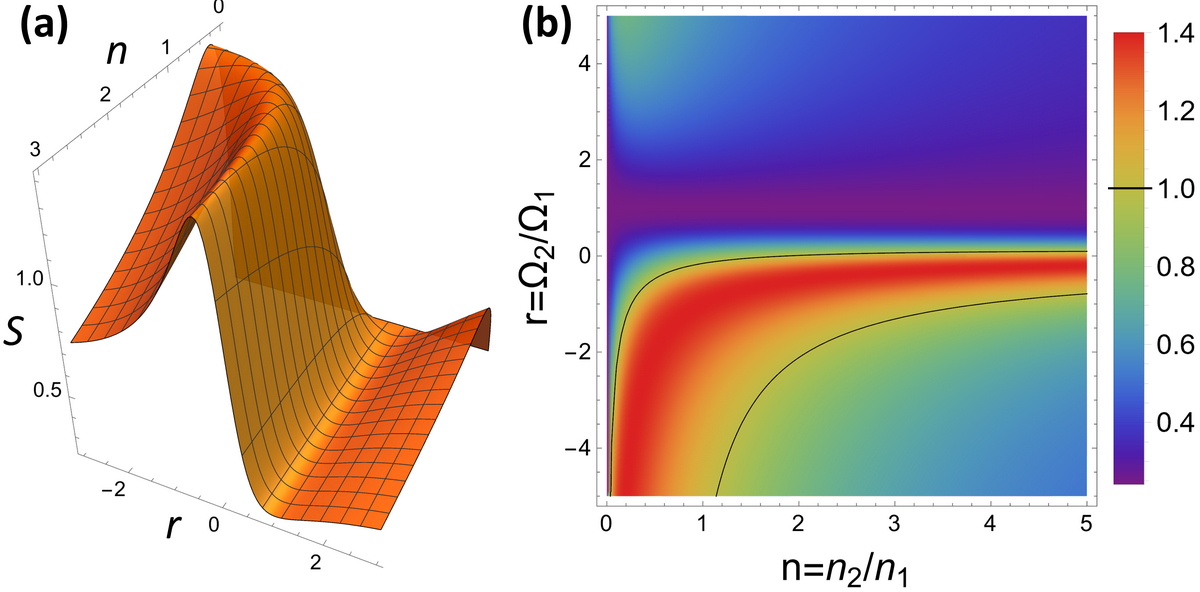}\caption{The slope of $H_{c2}$ at $t_{c}$, denoted here as $S\equiv\partial H_{c2}/\partial T|_{T=T_{c}}\equiv\partial_{T}H_{c2}|_{T_{c}}$,
plotted as a function of the ratio of the partial densities of states
$n=n_{2}/n_{1}$ and of the ratio of the angular parts of the order
parameter $r=\Omega_{2}/\Omega_{1}$. The region of $r\geq0$ corresponds
to $s_{++},$ whereas region $r<0$ to the $s_{\pm}$ pairing. Clearly, the slope $S$ behaves very differently in these two domains. }
\label{fig2:slope}
\end{figure}

\noindent We proceed in a similar way expressing various averages
in Eq.\ref{eq:slope-general} for the slope of $H_{c2}$ at $T_{c}$.
Introducing the ratio, $v=v_{a2}/v_{a1}$ and assuming a simple cylindrical
Fermi surface where $v_{a1}=v_{F1}\cos\varphi$, $v_{F1}$ is the
Fermi velocity on band one, we have:

\begin{equation}
\ensuremath{\left\langle \Omega\right\rangle =\frac{1+nr}{\sqrt{\left(1+n\right)\left(1+nr^{2}\right)}}}\label{eq:om}
\end{equation}

\begin{equation}
\ensuremath{\langle\Omega^{2}v_{a}^{2}\rangle=\frac{v_{F1}^{2}}{2}\frac{1+nr^{2}v^{2}}{1+nr^{2}}}\label{eq:om2v2}
\end{equation}

\begin{equation}
\ensuremath{\left\langle \Omega v_{a}^{2}\right\rangle =\frac{v_{F1}^{2}}{2}\frac{1+nrv^{2}}{\sqrt{\left(1+n\right)\left(1+nr^{2}\right)}}}\label{eq:omv2}
\end{equation}

\begin{equation}
\ensuremath{\left\langle v_{a}^{2}\right\rangle =\frac{v_{F1}^{2}}{2}\frac{1+nv^{2}}{1+n}}\label{eq:v2}
\end{equation}

\noindent where $1/2$ comes from $\left\langle \cos^{2}\varphi\right\rangle =1/2$
since we consider constant $\Omega_{i}$ and only the $\cos^{2}\varphi$ needs to be averaged. These equations are substituted into the general Eq.\ref{eq:slope-general} along with the coefficients, Eq.\ref{eq:h-i-j}, and the slope can be numerically evaluated for any values of $n,r,$ and $v$, characterizing the two-band superconductor, and for any
non-magnetic scattering rate $P$. One needs to watch for the meaningful range of the scattering parameter by ensuring that the transition temperature, Eq.\ref{eq:tc}, remains finite. For example, in a pure
$s_{\pm}$ situation, $n=1,\;v=1,\;r=-1$, $T_{c}$ is suppressed to zero at $P=0.2808,$ the same as in the $d-$wave order parameter, because in both cases, $\left\langle \Omega\right\rangle =0$.

\begin{figure}[b]
\includegraphics[width=8.5cm]{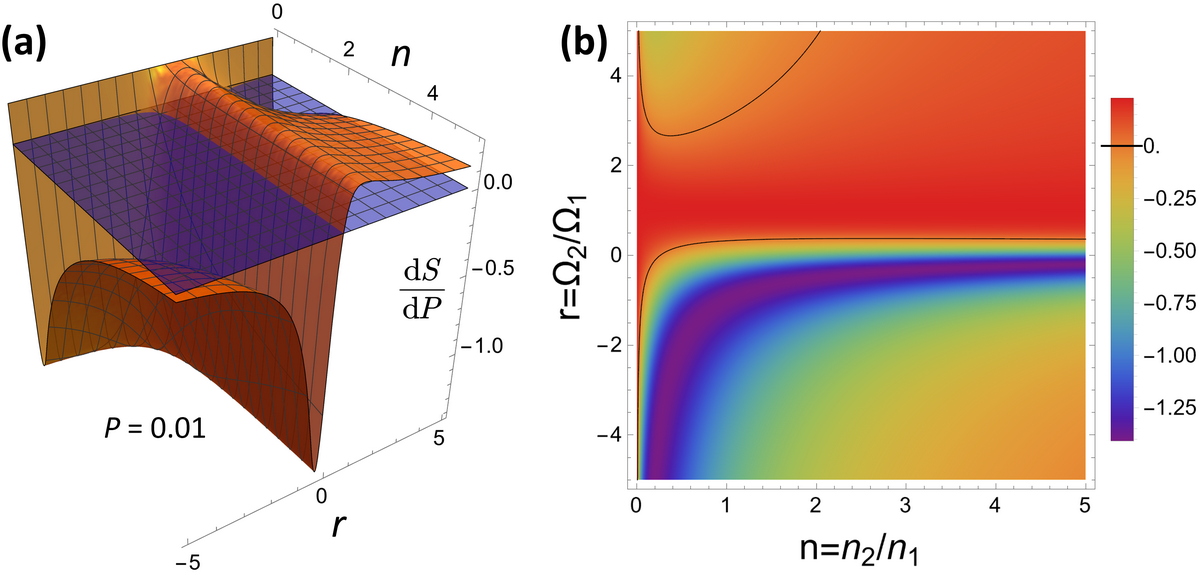}
\caption{Numerical derivative of the slope, use, $dS/dP\equiv[S(P=0.011)-S(P=0.010)]/0.001$, as function of $n$ and $r$. Negative values indicate slope decreasing with $P,$positive - increasing. Two contour lines on the density plot show $S=0$.}
\label{fig3:dSdP}
\end{figure}

Let us now examine some numerical solutions to the above equations. Figure \ref{fig1:tc} shows the superconducting transition temperature,
$t_{c}$, at the critical value, $P=0.2808$, Eq.\ref{eq:tc}, varying the gap ratio, $r$, and the ratio of the partial densities of states, $n$. The ratio of Fermi velocities was found not to affect results in a meaningful qualitative way; it only changes the amplitudes, so
we set it to $v=1$. A clearly asymmetric 3D surface shows a deep trench at negative $r$ (gaps of the opposite signs), corresponding to a suppression of $t_{c}$ to zero. This is even better seen on
the density plot, right panel of Fig.\ref{fig1:tc}. The two black contour lines inside outline the location of $t_{c}=0.05$ (one cannot set it to zero due to singularity in the derivative). The true $t_{c}=0$ line is located in between. By definition, $t_{c}$ is suppressed to zero when $\left\langle \Omega\right\rangle =0$ and
from Eq.\ref{eq:om}, we see that this line in the $n-r$ plane is $nr+1=0$, which is exactly what we have in Fig.\ref{fig1:tc}. Except for this line, all positive values of $n,$and both positive and negative values of $r$ are possible. Of course, in realistic iron-based superconductors, the ratio of the two effective gaps (supported by five Fermi surface pockets) is about two \cite{Prozorov2011RPP} (or, equivalently for
our model, 1/2). For positive $r$, expectedly, $t_{c}$ does not change much but decreases for larger values since the anisotropy increases. When we discuss the slope of $H_{c2}$, we need to consider what range of $P$ values makes sense. From Fig.\ref{fig1:tc}
is obvious that for $n$ in the interval between $0$ and $1$, maximum $P$ is not much larger than the critical value. Instead of a sharp termination, there is a long tail, but $t_{c}$ is already practically zero. For larger $n$, a large range of scattering parameters is valid.

\begin{figure}[tb]
\includegraphics[width=8.5cm]{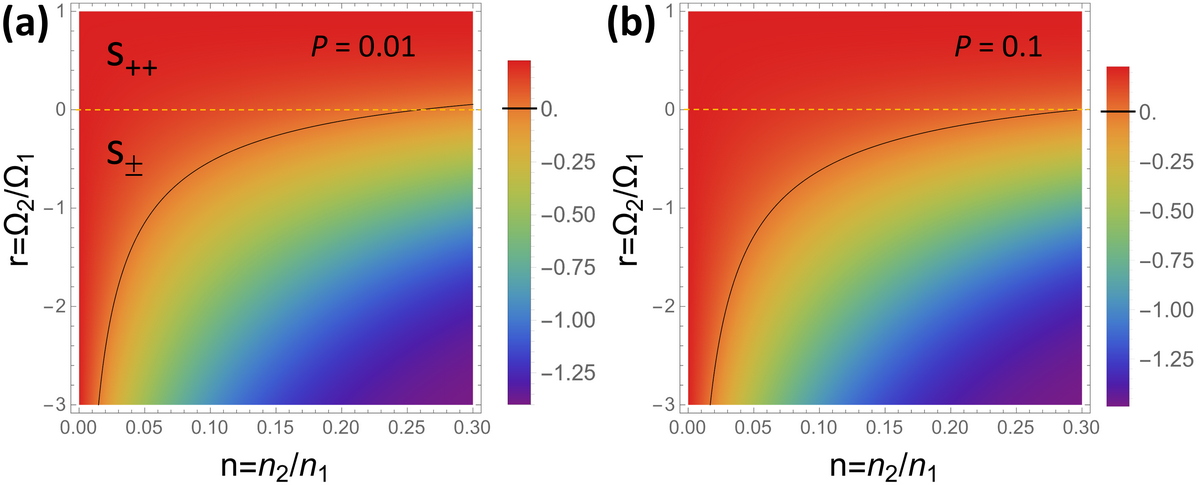}\caption{Two numerical derivatives of the slope. The left panel is the same as
in Fig.\ref{fig3:dSdP} (clean limit), but focusing on the region of small $n$. Right panel shows $dS/dP\equiv[S(P=0.101)-S(P=0.101)]/0.001,$
corresponding to a significant scattering,$P=0.1$ (compared to the critical value of $P=0.2808$). In the narrow region indicated by the red color, an increasing with $P$ slope exists in the $s_{\pm}$
pairing region.}
\label{fig4:dSdPcleanDirty}
\end{figure}

Now we can discuss the slope of $H_{c2}$ at $t_{c}$, denoted here as $S\equiv\partial H_{c2}/\partial T|_{T=T_{c}}\equiv\partial_{T}H_{c2}|_{T_{c}}$.
Figure \ref{fig2:slope} shows 3D and 2D plots of the slope $S(P=0.001,n,r,v=1)$ in the clean limit. A fairly complicated surface reveals significant
asymmetry with respect to positive and negative values of $r$. Two contour lines in the right panel show the location of $S=1.$ According to the color legend, the red domain in between is where $S>1$ and is less around it. However, the magnitude of slope $S$ does not
tell us whether it increases or decreases with scattering $P$. To probe the disorder dependence of $S$, we construct the numerical derivative for two values of $P$. In the clean limit we use, $dS/dP\equiv[S(P=0.011)-S(P=0.010)]/0.001$, and plot this quantity
in Fig.\ref{fig3:dSdP}. Here positive values indicate the increase of $S$ with the increase of $P$. Surprisingly, the result is non-trivial, showing different trends depending on $n$ and $r$.
A 2D density plot in Fig.\ref{fig3:dSdP} shows two black contour lines of $S=0$, indicating a large positive domain (in red) for positive $r$ values. Above roughly $r=3$, for reasonable $n\sim1$, the high anisotropy takes over even in this $s_{++}$ state, and the slope $S$ becomes a decreasing function of $P$. We note that we have also
explored the influence of the ratio of the Fermi velocities, $v$, but did not find much effect on the results.

\begin{figure}[t]
\includegraphics[width=8.0cm]{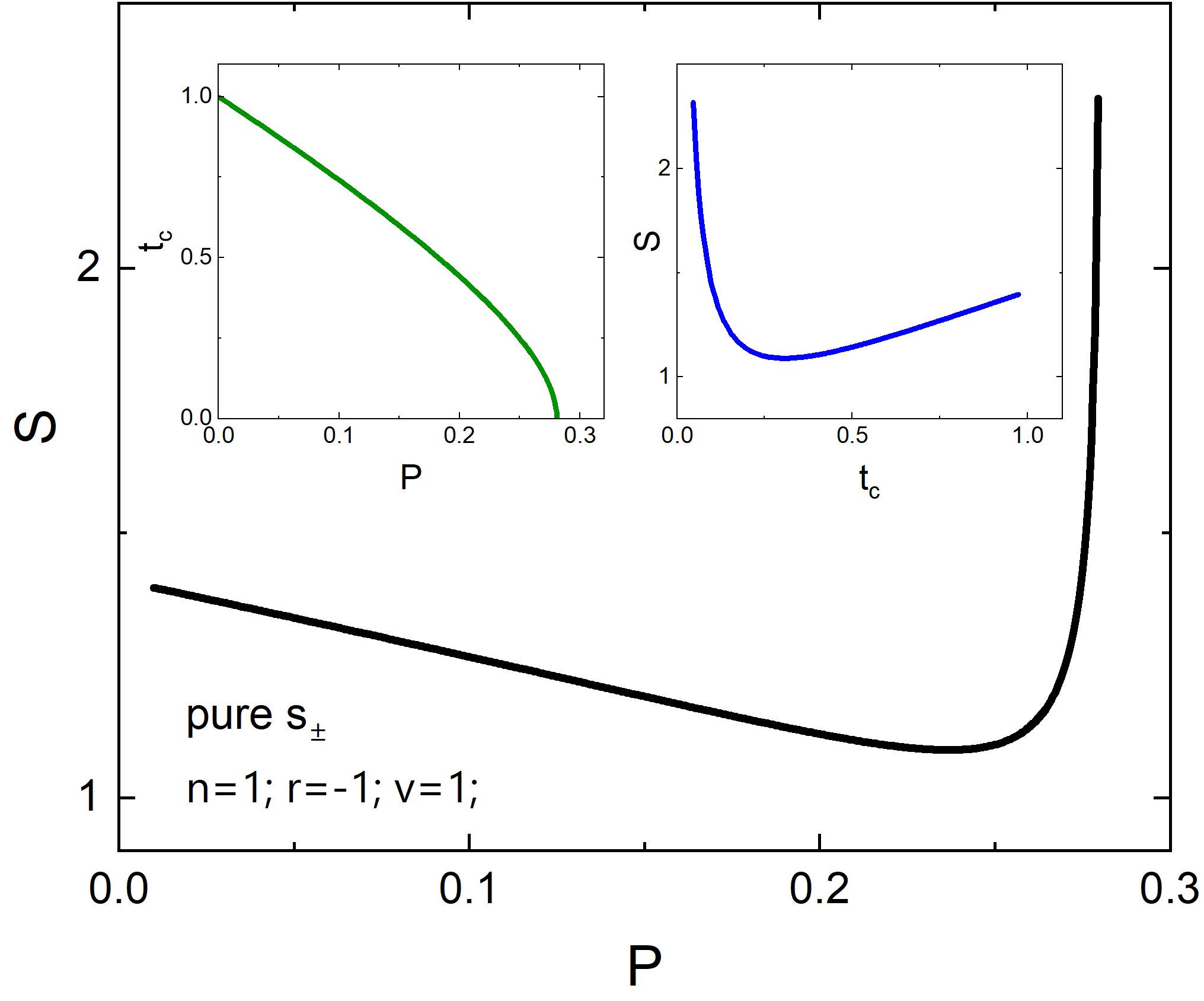}\caption{The slope of the upper critical field for Mazin's \cite{Mazin2010}
$s_{\pm}$ state, $n=1,$ $v=1,$ $r=-1$. This is identical to a superconductor with a $d-$wave order parameter; see Fig.3 in Ref.\cite{Kogan2023}. }
\label{fig5:purespm}
\end{figure}

Considering our experimental results, it is, however, more interesting to explore possible slope increase on the negative, $s_{\pm}$, side of $\text{r}$ values. Then, the only region of interest is at small $n$. Figure~\ref{fig4:dSdPcleanDirty} zooms on this region and, in addition to the clean limit $dS/dP$, shows a dirty-limit derivative, $dS/dP\equiv[S(P=0.101)-S(P=0.101)]/0.001$, offset by $P=0.1$ (about
third of the critical value of 0.28). According to the color legend, the red domain is where the slope of $H_{c2}$ at $T_{c}$ increases with the increase of non-magnetic scattering. This is only possible for practically non-physical values of the DOS ratio, $n<0.2$. In iron
pnictides, and more specifically, $\textrm{Ba}_{1-x}\textrm{K}_{x}\textrm{Fe}_{2}\textrm{As}_{2}$, which are the experimental subject of this paper, this ratio is not too large or too low. Furthermore, with a fairly two-dimensional character of the bands, the densities of states do not depend (much) on energy; therefore, doping while
shifting the Fermi level does not alter much the $n$ value \cite{Yin_2011}. As for the gaps ratio, $r-$values, experimentally it was found that $\left|r\right|\approx2$
(or, which is the same in our model,  $\left|r\right|\approx1/2$)
\cite{Mou2016,Ding2008}. Therefore, our model predicts that if iron pnictides are $s_{\pm}$ superconductors, the slope at $T_{c}$ of their upper critical field should decrease with the increasing
transport (non-magnetic) scattering.

\begin{figure}[tb]
\includegraphics[width=8.5cm]{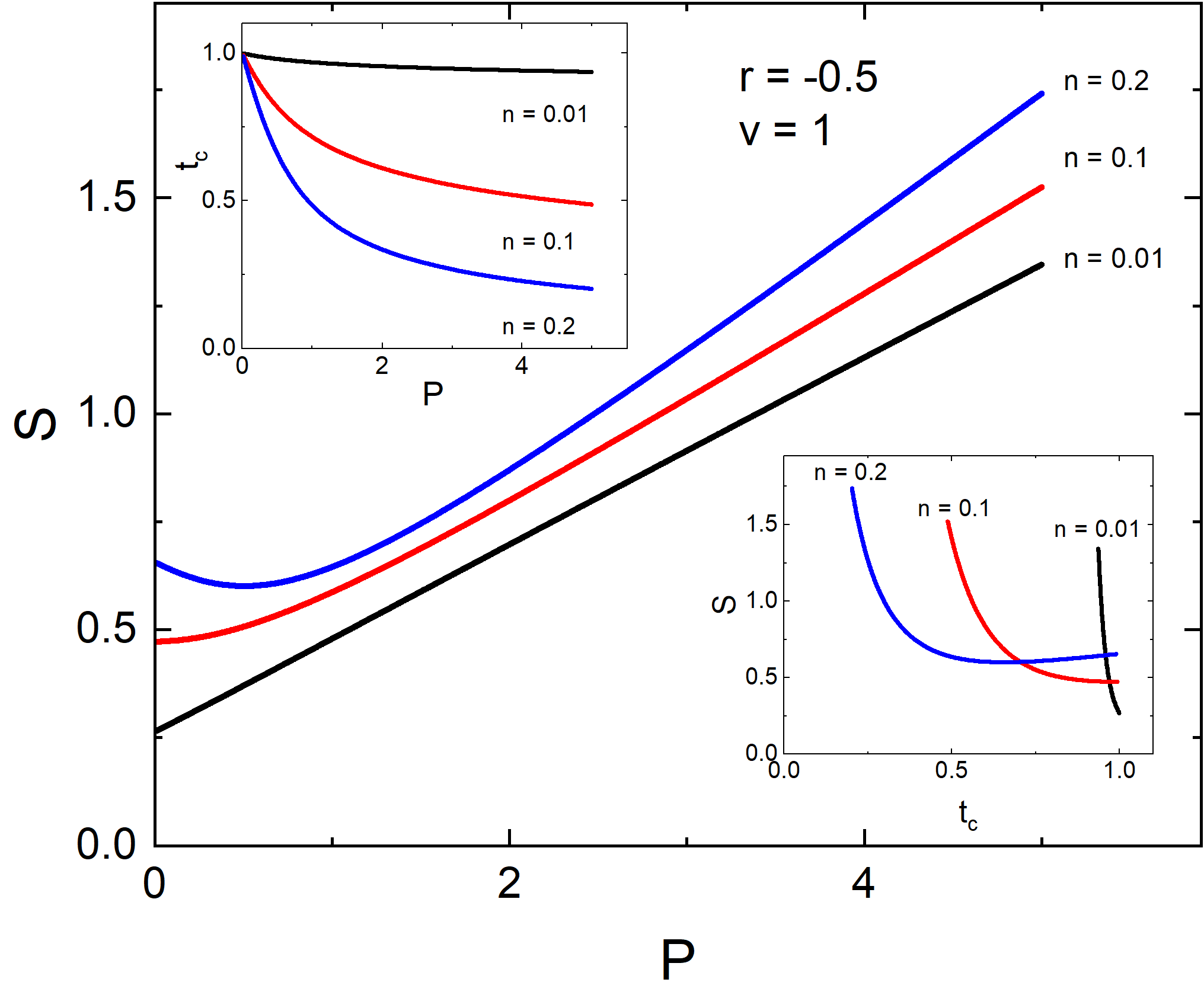}\caption{The slope $S$ in the narrow domain of very small $n$ where it shows
an increase with $P$ but almost immediately changes to a decreasing
function. The inset shows a corresponding reduction of the transition temperature
for the same $n$ values.}
\label{fig6:S-small-n}
\end{figure}

More explicitly, the expectations for a pure $s_{\pm}$ state are completely equivalent to a line nodal $d-$wave superconductor \cite{Kogan2023}. Figure \ref{fig5:purespm} shows the slope $S(P)$ for pure $s_{\pm}$ state where $n=1,v=1,$and $r=-1$. As in a $d-$wave, the $T_{c}$ is suppressed to zero at the critical value, $P=0.2808$ (top left inset), at which the slope sharply diverges. As discussed in Ref. \cite{Kogan2023}, there is a small interval where this state is gapless. Plotting as a function of $P$ is convenient for a theory; however, in practice, one would use the apparent (observed) transition temperature as the measure of the scattering rate. This is shown in
the upper right inset. The slope is predicted to decrease for most of the $T_{c}$ values.

Let us examine in more detail the region of $s_{\pm}$ side where the slope can increase, as discussed above, at $n<0.2$. Figure \ref{fig6:S-small-n}
shows thee curves taken at the fixed (experimental) $r=-0.5$ for $n=0.01,0.1,$and $0.2$. Already at $n=0.2,$ the slope starts from the decreasing trend. The suppression of $t_{c}$ is shown in the top left inset, whereas the slope vs. $t_{c}$ is shown in the
bottom right inset. Such behavior requires a very special set of electronic bandstructure parameters and cannot be robust and generic for a large family of compounds.

\begin{figure}[tb]
\includegraphics[width=8.5cm]{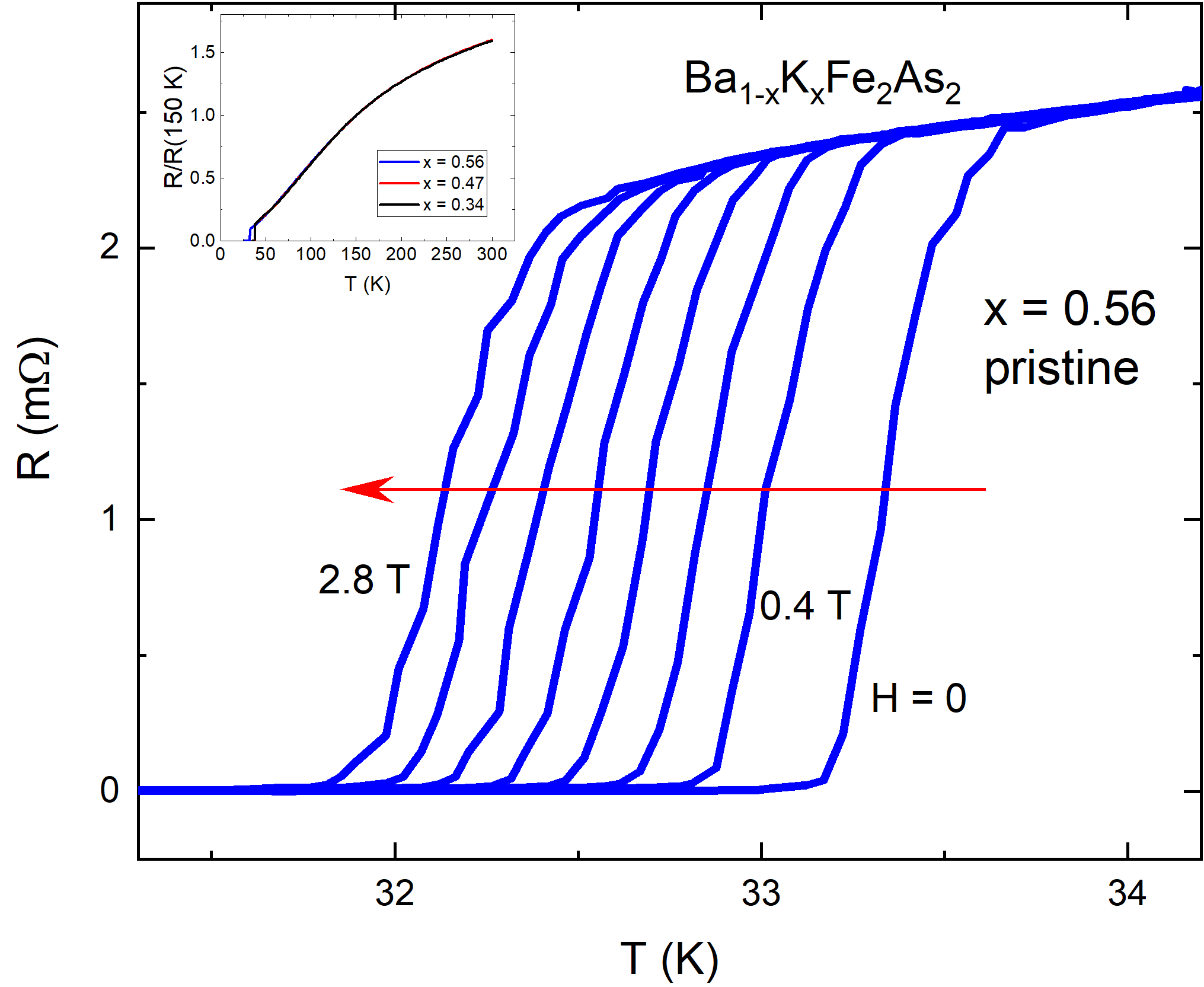}\caption{Temperature dependent resistance of pristine sample of BaK122, $x=0.56$.
Data are taken in magnetic fields applied along tetragonal $c-$axis from 0~T (right curve) to 2.8~T (left curve). Note the nearly parallel shift of the curves, which makes the same slope of the $H_{c2}(T)$ curve irrespective of the criterion used. Insert shows temperature
dependence of normalized resistance $R/R(300K)$ for compositions $x=$0.34 (black), $x=$0.47 (red) and $x=$0.56. }
\label{fig7:R(T)}
\end{figure}

\section{Experimental slope, $\bm{\partial H_{c2}/\partial T|_{T=T_{c}}}$,
in $\textrm{Ba}_{1-x}\textrm{K}_{x}\textrm{Fe}_{2}\textrm{As}_{2}$}

As a specific system to probe our theoretical conclusions, we selected a well-studied (Ba$_{1-x}$K$_{x}$)Fe$_{2}$As$_{2}$ family of iron-based
superconductors (abbreviated as BaK122). Considering the significant dependence of the results on $n$, it was important to probe several different compositions. An example of the data collected in overdoped
BaK122, $x=0.56$, is shown in Fig.\ref{fig7:R(T)} where temperature-dependent resistance is plotted for several values of the applied magnetic field. The inset shows a full temperature dependence of the resistance normalized by the room temperature value. The curves are parallel and not smeared, allowing us to use an easy criterion of 50\%  of the transition to estimate $H_{c2}$. The actual values are not important,
the important is the functional dependence of the slope on the scattering parameter.

\begin{figure}[tb]
\includegraphics[width=8.5cm]{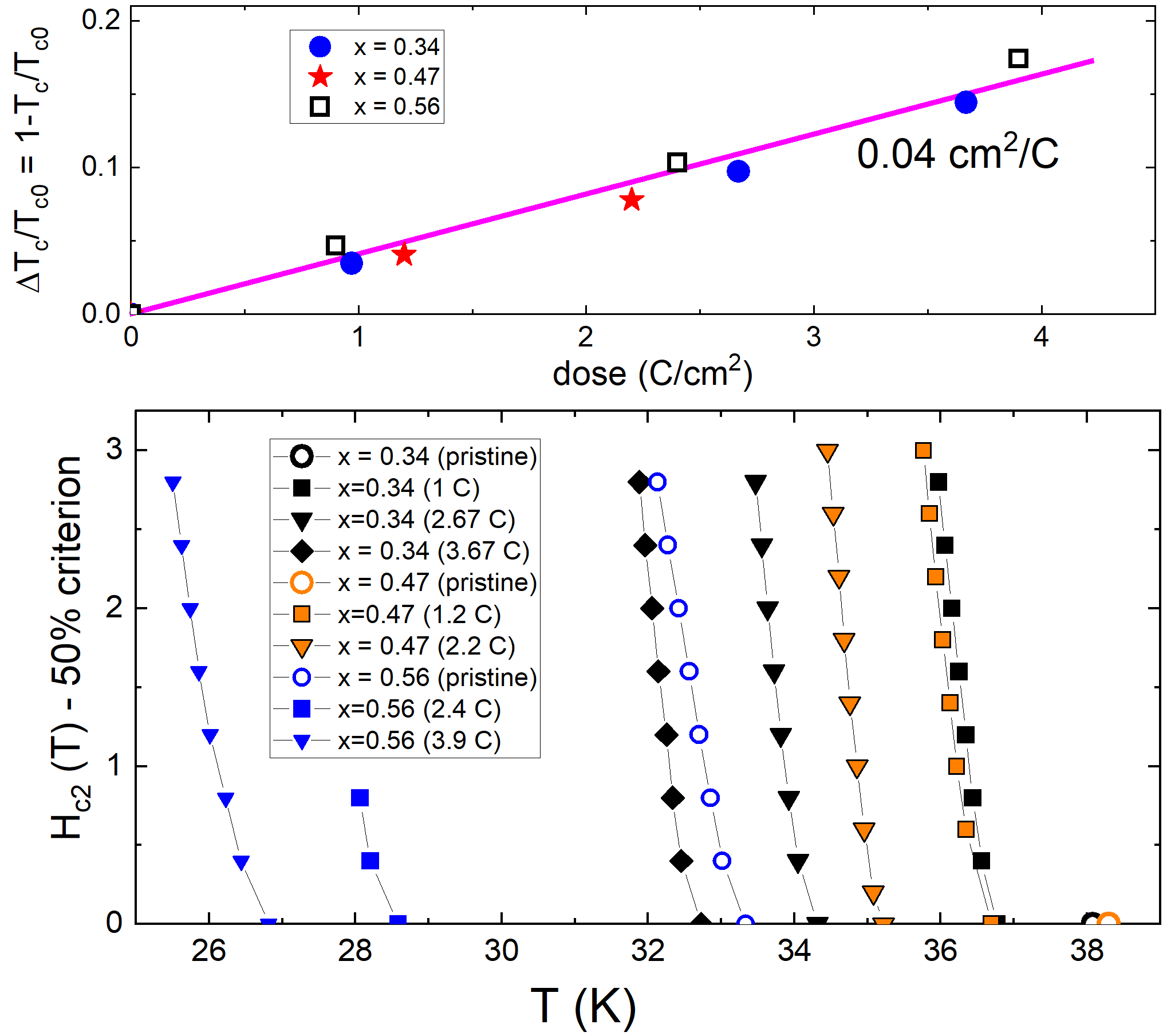}\caption{(Top panel) Normalized variation of the transition temperature, $\varDelta T_{c}/T_{c0}$ plotted versus the irradiation dose showing a universal behavior for all studied compositions because it depends only on the scattering
rate $P$, which is a linear function of the dose. (Bottom panel) The upper critical field as a function of temperature near $T_{c}$ for select compositions is shown in the legend. }
\label{fig8:Hc2}
\end{figure}

Four different compositions were measured, $x=0.2$ (under doped), $x=0.34$ (optimally doped), and two moderately overdoped, $x=0.47$ and $x=0.56$, but before the Lifshitz transition, which alters
the electronic band structure and the gap structure considerably \cite{Cho2016}. Each sample was measured as shown in Fig.\ref{fig7:R(T)}, then put into the irradiation chamber, irradiated with the dose shown, extracted and brought to room temperature, then measured again, and so forth, the cycle repeated. The top panel in Fig.\ref{fig8:Hc2} shows the change of the superconducting transition
temperature after consequent irradiations. These results are expected. As shown before, the dimensionless scattering rate induced by electron irradiation is linearly proportional to the dose, at least for relatively small doses, and $T_{c}$ is also linearly suppressed
with $P$. We remind that here the suppression of $T_{c}$ is only by non-magnetic defects, therefore only due to the anisotropy of the order parameter described by the $\Omega$ functions of our two-band
system, Eq.\ref{eq:om}.

\begin{figure}[tb]
\includegraphics[width=8.5cm]{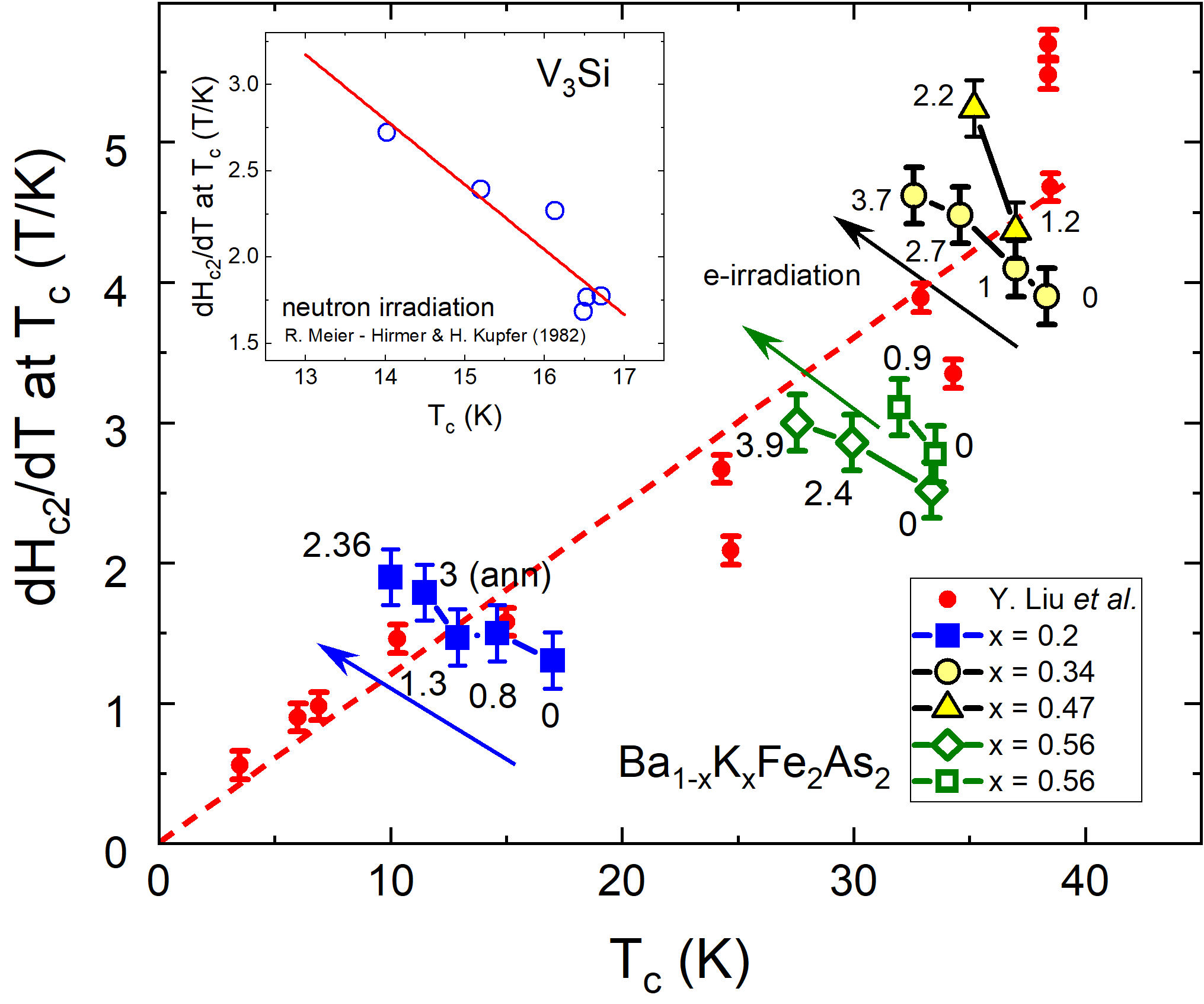}\caption{The experimental slope of the upper critical field. Red symbols show many different compositions in their pristine state. As expected from BCS theory, the slope is proportional to $T_{c}.$ Blue, green and
yellow symbols show the slope change with electron irradiation for compositions indicated in the legend. The numbers next to symbols are the doses in C/cm$^2$. Inset shows a quite similar behavior in a known two-band $s_{++}$ superconductor V$_{3}$Si \cite{Cho2022}  irradiated
by neutrons \cite{Meier-Hirmer1982}.}
\label{fig9:S-vs-Tc}
\end{figure}

The lower panel of \ref{fig8:Hc2} shows measured $H_{c2}$ before and after irradiating the indicated compounds. The slope was evaluated as a linear derivative of each curve. The summary of the results is presented in Fig.\ref{fig9:S-vs-Tc}. Blue, green and
yellow symbols show the slope change with electron irradiation for compositions indicated in the legend. The numbers next to symbols are the doses in C/cm$^2$. The slope $S$ increases with decreasing transition temperature, proportional to the scattering rate, $P$. For comparison, the inset in Fig.\ref{fig9:S-vs-Tc} shows similar data collected on a known two-band $s_{++}$ superconductor, V$_{3}$Si \cite{Cho2022}, irradiated by neutrons \cite{Meier-Hirmer1982}. The slope $S$ increases as expected from our model. This behavior is contrasted with the red symbols (main panel) showing the slope $S$ as a function of $T_{c}$ in pristine
compositions of $\textrm{Ba}_{1-x}\textrm{K}_{x}\textrm{Fe}_{2}\textrm{As}_{2}$, revealing expected from the BCS linear proportionality, $S \propto T_c$  \cite{Kogan2023}.

In another experiment on a single crystal of (Ba$_{0.75}$K$_{0.25}$)Fe$_{2}$As$_{2}$
($T_{c0}\approx30.3$ K), the 2.5 MeV electron irradiation was pushed to a very large dose of $\ensuremath{\text{8.93 C/cm}^{2}=5.6\times10^{19}\text{electrons/cm}^{2}}$.
To put this in perspective, a usual overnight irradiation run yields around 0.8 C/cm$^{2}$, so $\text{8.93 C/cm}$ would be achieved in about five days of continuous irradiation, which is impossible to do in one go. It took about two weeks of active irradiation spread over
several sessions that lasted a few years.

\begin{figure}[tb]
\includegraphics[width=8.5cm]{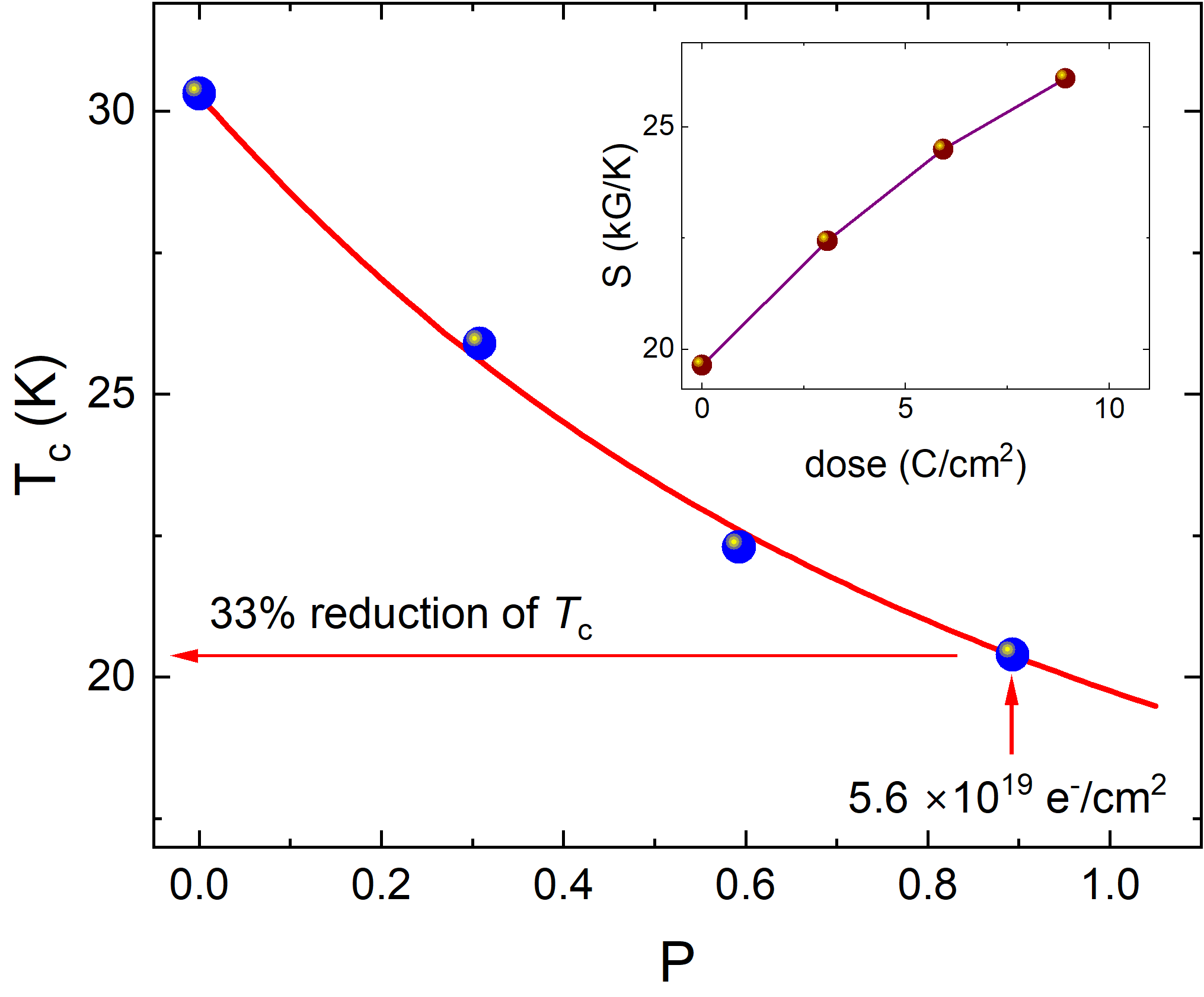}\caption{Superconducting transition temperature, $T_{c}$, versus non-magnetic
scattering rate, $P$. Symbols are the experimental values obtained
on nearly optimally-doped $\textrm{Ba}_{1-x}\textrm{K}_{x}\textrm{Fe}_{2}\textrm{As}_{2}$
crystal irradiated with the doses of 3.08 C/cm$^{2}$ ($\ensuremath{1.92\times10^{19}}$
electrons/cm$^{2}$), 5.93 C/cm$^{2}$ ($\ensuremath{3.70\times10^{19}}$
electrons/cm$^{2}$), and $\ensuremath{\text{8.93 C/cm}^{2}}$($5.57\times10^{19}\text{electrons/cm}^{2}$).
The solid red line is a fit to Eq.\ref{fig1:tc}, with $s_{++}$ pairing parameters, $n=0.3$ and $r=2.95$ with $v=1$ kept constant. Inset shows the slope $S$ as function of $P$.}
\label{fig9:Tc-vs-P}
\end{figure}

In BaFe$_{2}$As$_{2}$, thresholds energies of ion knockout upon head-on collisions, $E_{d}$, were calculated using VASP-MD simulation
that yielded $E_{d}=$33 eV (Ba), 22 eV (Fe), and 50 eV (As) \cite{Cho2023}. With these numbers, we used SECTE software (for details, see Ref.\cite{Cho2023})
to calculate the total cross-section of defects production upon electron irradiation, $\sigma=80$ barn at 2.5 MeV. This gives $5\times10^{-4}$
dpa (defects per atom) per 1 C/cm$^{2}$. For our largest dose of $\text{8.93 C/cm}^{2}$ we estimate $4.5\times10^{-3}$ dpa or $0.045$
defects per a conventional unit cell ($Z=2$). This means that we produce one defect per $22.2$ conventional unit cells at this dose. Therefore, with the unit cell volume of $0.20457$ nm$^{3}$, the average distance between the defects is $1.66$ nm. This should be compared to the coherence length, $\xi$, and Bardeen-Cooper-Schrieffer (BCS) coherence length, $\xi_{0}=\hbar v_{F}/\pi\varDelta_{0}$ \cite{TinkhamBOOK}.
Ba$_{1-x}$K$_{x}$Fe$_{2}$As$_{2}$ at the optimal doping, $x=0.4$,
$T_{c}=38$ K, has $H_{c2}$ with $H||c-$axis of about 150 T, whereas
our somewhat underdoped sample has $H_{c2}=70\:\mathrm{T}$ \cite{LiuProzorovLograsso2014PRB}.
Therefore while the optimal composition would have $\xi=\sqrt{\phi_{0}/2\pi H_{c2}}\approx1.5$ nm, our underdoped sample gives $\xi=2.2$ nm, both comparable with the estimated
inter-defect distance. Away from the optimal doping, the upper critical
field and transition temperature, $T_{c}$, decrease substantially,
which means that those compositions will be deeper in the dirty limit
since the scattering rate $P\propto\xi_{0}/\ell$, where $\ell$ is
the mean free path, $\ell\propto\mathrm{dpa}$. With $\hbar v_{F}\sim0.7\:\mathrm{eV}\text{\AA}$,
the BCS coherence length, $\xi_{0}\approx209.88\hbar v_{F}\left[\mathrm{eV}\text{\AA}\right]/T_{c}\left[K\right]$,
is about 4 nm at the optimal doping and about 5 nm for our $x=0.25.$ Therefore, in our particular study, we expect $P\lesssim1$ for all irradiation doses, which is precisely what we obtain in Fig.\ref{fig9:Tc-vs-P}
that shows the experimental superconducting transition temperature,
$T_{c}$, as a function of the scattering rate $P$ (symbols) and
the fit to Eq.\ref{fig1:tc} with $\Omega$ described by $s_{++}$
parameters, $n=0.3$ and $r=+2.95$ with $v=1$ (kept constant). Due
to high irradiation dose, the $T_{c}$ decreased substantially, by
33\% at the largest dose of $\ensuremath{\text{8.93 C/cm}^{2}}$($5.57\times10^{19}\text{electrons/cm}^{2}$).
The intermediate doses were, 3.08 C/cm$^{2}$ ($\ensuremath{1.92\times10^{19}}$
electrons/cm$^{2}$) and 5.93 C/cm$^{2}$ ($\ensuremath{3.70\times10^{19}}$
electrons/cm$^{2}$). The inset in Fig.\ref{fig9:Tc-vs-P} shows the
slope $S$ increasing with the scattering rate, $P$, as expected
for an $s_{++}$ pairing from the above theory.

\section{Discussion}

The results of the theoretical part are straightforward. Unless a superconductor has very imbalanced partial densities of states of the order of 10\% or less, it will show a reduction of the slope of $H_{c2}$ at $T_{c}$ with transport (non-magnetic) disorder if it has nodes or any other state in which the order parameter has different signs, such as $s_{\pm}$ multiband superconductivity. An increasing slope, including an anisotropic multiband case, is predicted for any $s_{++}$ state. Note, however, that the initial change in the slope becomes negative at a very large difference between gap amplitudes and small $n$; see the upper left corner of the right panel in Fig.\ref{fig3:dSdP}.

Our data in $\textrm{Ba}_{1-x}\textrm{K}_{x}\textrm{Fe}_{2}\textrm{As}_{2}$ show that the slope $S$ increases with $P$ across the $T_{c}\left(x\right)$ dome of superconductivity for underdoped, optimally doped and overdoped compositions. This is a strong argument
in favor of a multiband $s_{++}$ superconductivity with a significant difference between different gaps. Reviewing the literature and our work of the past two decades, there are no experimental facts, at least for $\textrm{Ba}_{1-x}\textrm{K}_{x}\textrm{Fe}_{2}\textrm{As}_{2}$, that could not be explained from the anisotropic $s_{++}$ point of view. This includes the suppression of $T_{c}$, non-exponential London penetration depth, specific heat, thermal conductivity, and other transport and thermodynamic quantities. Angle-resolved photo-emission is not sensitive to the sign of the order parameter but produced important information regarding
the gaps anisotropy on separate sheets of the Fermi surface \cite{Borisenko2010,Ding2008,Okazaki2012,Umezawa2012}. Importantly, most works find a fully gapped robust superconductivity in BaK122 except for the end member, KFe$_{2}$As$_{2}$, which is nodal \cite{Reid2012,Reid2012a}. The only phase-sensitive experiment that directly confirmed $s_{\pm}$ superconductivity, quasiparticle interference, was only successfully performed on chalcogenides \cite{Hanaguri2010,Sprau2017},
which are remote cousins of the pnictides.

Of course, the question of the pairing type is complicated and requires considering multiple independent experiments and theories. For example,
tunneling spectroscopy and neutron resonance studies \cite{Scalapino2012} bring important information linked directly to the nature of the interactions in the system. While our approach is based on a general Ginzburg-Landau treatment of the upper critical field at $T_{c}$, our two-band model is simple. Perhaps, a more elaborate microscopic theory would find something else. However, the obtained results are practically expected. It is the experiment that showed the trend opposite to what is predicted for an $s_{\pm}$ superconductor. Do we have a smoking gun proving without a doubt the $s_{++}$pairing in $\textrm{Ba}_{1-x}\textrm{K}_{x}\textrm{Fe}_{2}\textrm{As}_{2}$?
No, we do not, but what seems settled in the community is now re-opened for a more in-depth discussion.

\section{Methods}

Single crystals of Ba$_{0.2}$K$_{0.8}$Fe$_{2}$As$_{2}$ were grown by using an inverted temperate gradient method with the starting materials, Ba and K lumps, and Fe and As powders. Details of the growth are
published elsewhere \cite{Liu2013,LiuProzorovLograsso2014PRB,Cho2016}.
Resistivity measurements were performed in a standard four-probe configuration. Typical dimensions of the samples are (1-2) $\times$ 0.5 $\times$ (0.02-0.1) mm$^{3}$. Silver wires of 50 $\mu$m diameter were soldered to the sample to provide electrical contacts \cite{Tanatar2010SST}. The sample with four contact wires attached was mounted on a hollowed Kyocera chip (for the electron beam to propagate) over a hole of about 5 mm diameter in the center. After receiving a certain dose, the Kyocera chip was extracted and mounted in a different cryostat without disturbing the sample or the contacts. After resistance vs. temperature at different applied magnetic fields was measured, the Kyocera chip was returned to the irradiation chamber, and the process was repeated. The same procedure was performed on samples of different compositions.

The 2.5 MeV electron irradiation was performed at the SIRIUS Pelletron-type linear accelerator operating in the Laboratoire des Solides Irradi\'{e}s at the \'{E}cole Polytechnique in Palaiseau, France. The acquired irradiation dose is conveniently
measured in C/cm$^{2}$, where 1 C/cm$^{2} = 6.24 \times 10^{18}$ electrons/cm$^{2}$. A Faraday cup placed behind the sample chamber enabled accurate measurement of the acquired dose during irradiation. The electron irradiation was performed in liquid hydrogen at 20 K to prevent Frenkel pairs recombination and defects clustering. The typical concentration of the induced defects is one defect per thousand of unit cells. Here, our highest dose of 8.93 C/cm$^2$ corresponds to about one defect per 22 conventional unit cells (Z=2 for BaFe$_2$As$_2$). Details of the irradiation experiments are available elsewhere \cite{Damask1963,Thompson1969,Cho2018SST_review_e-irr}.

\section{acknowledgments}

We thank P. Hirschfeld, A. Chubukov and T. Hanaguri for useful discussions.
We thank Y. Liu and T. Lograsso for providing high quality single crystals
of Ba$_{1-x}$K$_{x}$Fe$_{2}$As$_{2}$. This work was supported by the U.S. Department of Energy
(DOE), Office of Science, Basic Energy Sciences, Materials Science
and Engineering Division. Ames National Laboratory is operated for the U.S.
DOE by Iowa State University under contract DE-AC02-07CH11358. We
thank the SIRIUS team, O. Cavani, B. Boizot, V. Metayer, and J. Losco,
for running electron irradiation at \'{E}cole Polytechnique, Palaiseau,
France. The irradiation was supported by EMIR\&A network, under user proposal 11-11-0121.

%\bibliographystyle{apsrev4-2}
%\bibliography{SlopeHc2}


\end{document}